\documentclass{svproc}
\usepackage{url}

\usepackage{placeins}
\usepackage{verbatim}       
\usepackage{xcolor}         
\usepackage[hidelinks]{hyperref}       
\usepackage{url}            
\usepackage{booktabs}       
\usepackage{amsfonts}       

\usepackage{mathtools}

\usepackage{subcaption}

\usepackage{graphicx}
\usepackage{tabularx}
\usepackage{adjustbox}
\usepackage{longtable}
\usepackage{placeins}
\usepackage{mathtools}

\usepackage{algorithm}
\usepackage[noend]{algpseudocode}

\newcommand{\etal}{\textit{et al.}}

\usepackage{multirow}

\usepackage{comment}
\usepackage{cancel}
\usepackage{ulem}

\begin{document}
\mainmatter              
\title{Effectiveness Of Using Remote Laboratory in Promoting Simulation and Verification Tools}

\titlerunning{Effectiveness Of Using Remote Laboratory}  
%
\author{Shuowei Li$^{[0000-0003-4934-4691]}$ \and Rania Hussein$^{[0000-0002-2859-9401]}$}
\authorrunning{S Li \etal} 
%
\tocauthor{XXX, XXX}
\institute{University of Washington, Seattle WA 98195, USA\\
\email{\{lis26, ruhssein\}@uw.edu}}

\maketitle              

\begin{abstract}
\label{abstract}
The transition to remote learning during the pandemic has necessitated the development of new methods for conducting hands-on experiments. One significant challenge in this transition has been providing students with reliable and sustainable access to necessary hardware components, particularly for courses that require substantial equipment. Additionally, industry partners have high expectations for students to be proficient in simulation and verification tools. To address these challenges, we implemented a virtual breadboard feature that allows students to remotely access Field Programmable Gate Array (FPGA) hardware and complete lab assignments. Our evaluation of this approach, which included surveys of students and industry partners, revealed that it effectively transformed a traditionally in-person lab assignment into an online modality. Furthermore, this paper presents the perspective of industry professionals on verification and simulation tools as a highly desirable skill in the industry, a skill that remote labs tend to emphasize which makes remote labs a viable educational solution that can continue to be utilized even after the pandemic.

\end{abstract}
\section{Introduction}
\label{intro}
Universities have historically taught Field Programmable Gate Arrays (FPGAs) courses utilizing on-campus labs, which are equipped with workbenches, PCs, and other auxiliary tools~\cite{9163773}. Students must be physically present in the laboratories to engage with this equipment and complete assignments. Due to the Covid-19 epidemic, a new teaching strategy focused on leveraging a worldwide network of remote laboratories was required because this technique was not practical. According to research, students who use remote lab facilities are better prepared to build engineering systems~\cite{hussein2021}.

The purpose of remote laboratory facilities is to remotely operate and manage specialized experimental facilities in order to teach students knowledge in many engineering domains. 
Traditionally, online education is not favored in engineering courses requiring students to be physically present in the laboratory, since hands-on physical experiences play a central role in engineering education~\cite{Cooper2005RemoteLI,Grober_2010}. In addition, hands-on physical experiments help students achieve the expected learning outcome, promote reasoning skills, and professional preparation for the workforce~\cite{Basey2008,Hofstein2004}.

With computer technology advancements, people have introduced multimedia to facilitate engineering education. Examples of multimedia include interactive on-screen experiments procures, virtual simulations, automated data acquisition, and quantitative analysis of experiments. These computer-based technologies redefined the "hands-on" experiences.
The Covid-19 pandemic further simulates the widespread proliferation of remote laboratories. Remote laboratories have gained popularity, particularly during the Covid-19 pandemic, due to their accessibility, ease of use, and affordability. As a result, various initiatives for creating remote and virtual laboratory systems have been put forth.

Speech-based virtual assistants have been applied to various application areas. Callaghan \etal~\cite{10.1007/978-3-319-95678-7_63} investigated how to use virtual reality, the Internet of Things (IoT), and voice-controlled virtual assistants to guide students through their experiments, provide supplemental teaching resources as needed, and access, control, and configure instruments and hardware while providing formative and summative feedback.
Ak \etal~\cite{Ak2018} presented the design and implementation of the remote laboratory and learning management system.
Monzo \etal~\cite{Monzo2021} presented the Remote Laboratory at the Open University of Catalonia (RLAB-UOC) that allows engineering students studying online to conduct practical experiments using advanced electronic and communication equipment anywhere and anytime. Furthermore, this work presented the case of a blended learning approach adopted in the remote teaching of electrical power engineering by Arefi \etal~\cite{9589652} in 2021. 

\begin{figure}[!htbp]
    \centering
    \includegraphics[width=\textwidth]{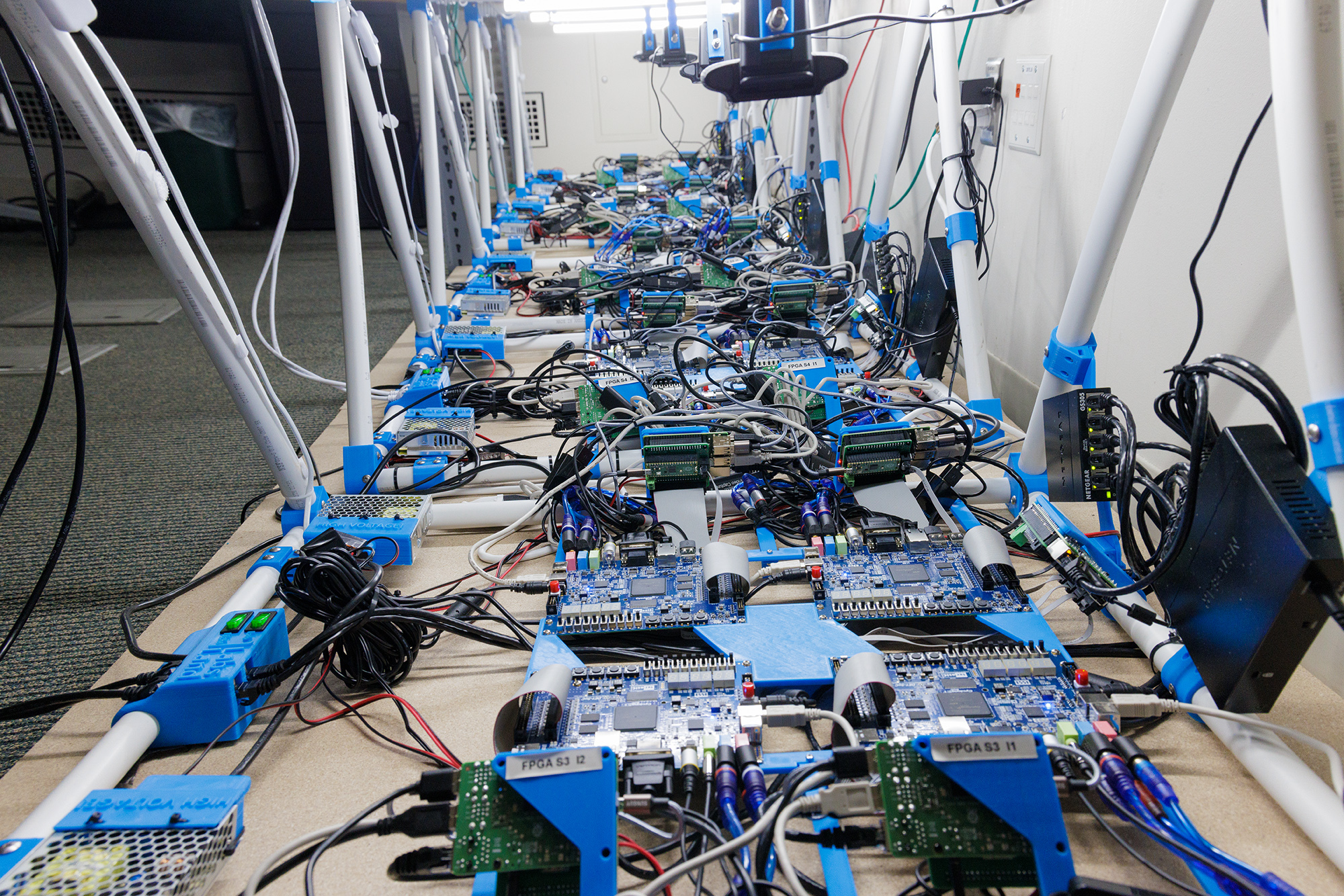}
    \caption{Remote FPGA lab in the Remote Hub Lab (RHLab).}
    \label{fig:remote_lab}
\end{figure}

Li \etal~\cite{10.1007/978-3-030-82529-4_15} introduces a feature that uses a virtual breadboard, where students can use GPIO ports to interface with physical FPGA boards that are accessed remotely through a global network called LabsLand. Figure~\ref{fig:remote_lab} shows the remote FGPA setup that is located in the Remote Hub Lab\footnote[1]{https://rhlab.ece.uw.edu}.
The proposed framework has been evaluated using a survey collected from students who enrolled in the class. The students' interviews focused on their experiences with the remote laboratory. In this paper, we extend the evaluation process to industry partners. The interviews with industry professionals zeroed in on their overall experience with simulation and verification tools. By comparing the results of the two surveys, we found that remote lab forces students to spend more time on simulation tools compared to traditional offline design. Therefore, students have considerably improved their skills in using simulation tools. This founding aligns with the study results of Hussein and Wilson’s work~\cite{hussein2021}.

\FloatBarrier
\section{Background}
\label{background}
\begin{figure}[!htbp]
    \centering
    \includegraphics[width=\textwidth]{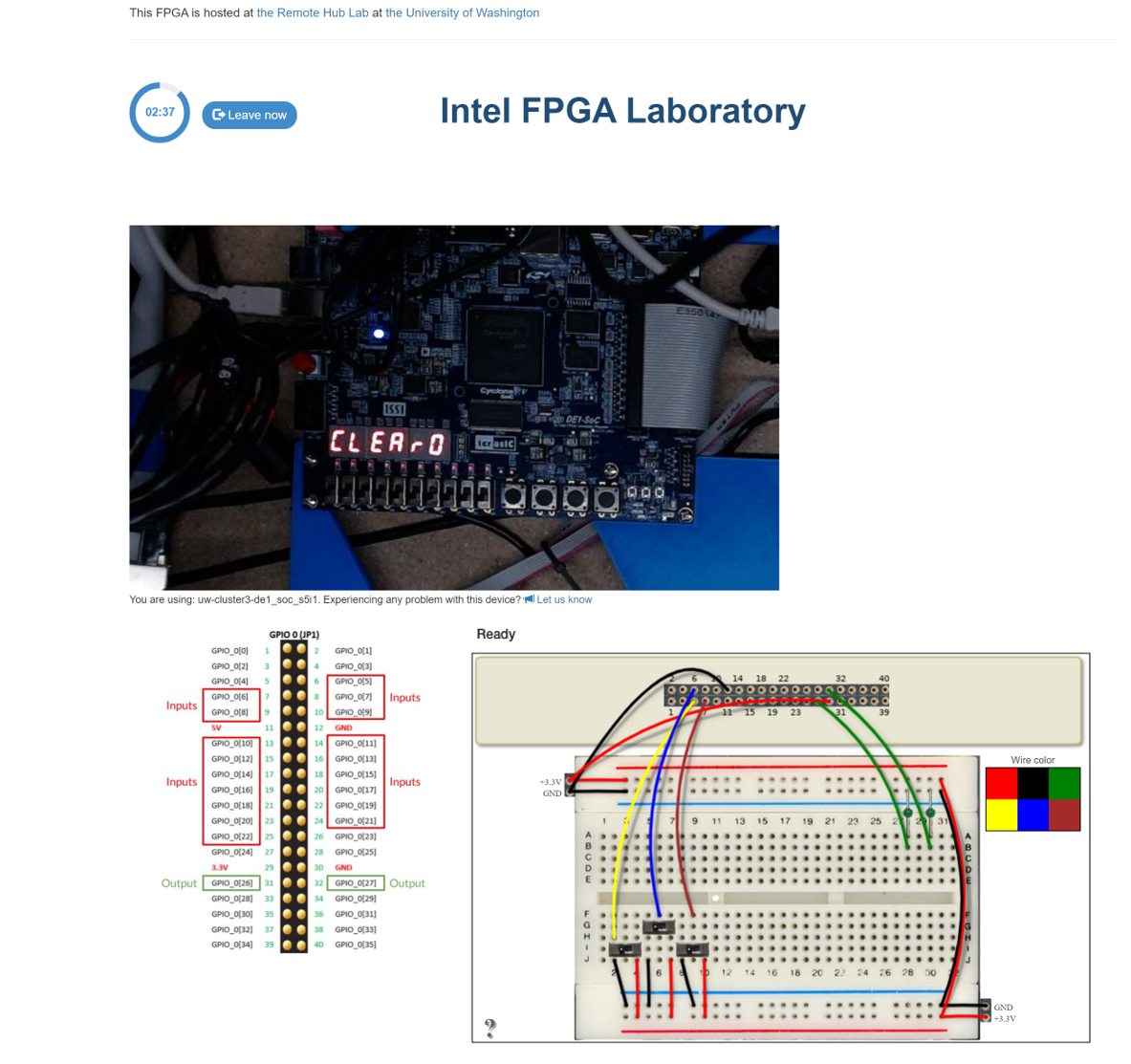}
    \caption{Intel FPGA Laboratory.}
    \label{fig:fpga_lab}
\end{figure}

Figure \ref{fig:fpga_lab} shows the students' view of the remote laboratory. The top figure shows a live video of FPGA with the compilation of students' programs. The virtual breadboard setup is demonstrated in the bottom figure.

\newpage
To give students an unrivaled learning experience, a virtual breadboard interface is integrated into the remote laboratory facility to emulate hands-on circuit assembly, allowing students to interface GPIOs in their SystemVerilog designs.

We place three switches and two LEDs on the virtual breadboard. Students have access to these circuit components via a configurator. A GPIO header is connected to the JP1 of the DE1-SoC board to allow the connections. As shown in the bottom left of Figure \ref{fig:fpga_lab}, a pinout diagram for the GPIO 0 (JP1) of the DE1-SoC is provided. This pinout diagram provides students with a more comprehensive view of the relationship between pin names and the construction of the circuit.

We ask students to connect each circuit component using the provided virtual breadboard, and then choose a wire to connect the circuit components. To simulate the real-world experience, all connections on the breadboard will be automatically preserved. Students can then proceed to the remote laboratory's web interface in the form of a code editor interface to finish implementing the built virtual breadboard circuit. To validate their design, students can flip the switches and see the states of the LEDs and other outputs on the DE1-SoC board, such as the HEX display. Students are given a tutorial for the breadboard user interface to help them become acquainted with the workflow.

Once the compilation is done, students can send the code to an actual FPGA that host at the UW campus to verify their designs. The top figure of Figure \ref{fig:fpga_lab} shows an example of the streaming image that are available to students.

\FloatBarrier
\section{Evaluate FPGA with Breadboard}
\label{experience}
The students in the Design Of Digital Circuits class at the UW in Autumn 2020 were asked to use the remote laboratory in addition to the newly designed breadboard user interface because of the logistical complexity of shipping physical DE1-SoC lab kits to students during the Covid-19 pandemic. The remote laboratory is accessible from anywhere at any time, and no laboratory equipment needs to be shipped. All students are required to complete a junior-level Design of Digital Circuits and System course before enrolling. Due to the recent deployment of the remote laboratory facilities for this framework, students only have prior experience with electronic components using physical breadboards and a DE1-SoC development board.

Li \etal \ evaluates the effectiveness of using a remote laboratory by users. According to the survey results, the virtual breadboard served its goal as a replacement for the physical breadboard. Furthermore, conscious design considerations, such as the positioning of the pinout diagram, aided students in understanding the course topics. Furthermore, despite closely resembling the original breadboard, the virtual breadboard is simple to use, allowing students to become adept quickly and save time on other activities. However, the poll results demonstrate that the breadboard interface has an opportunity for enhancement in the future.

We further extend our evaluation process to industry professionals who work in engineering-related fields. A new survey for a study on the effectiveness of simulation and experimentation tools in engineering design has been developed. In this survey, we seek the perspective of industry professionals who use modeling and simulation tools to some extent in their job duties. Industry professionals are asked to fill out the survey based on their collective work experience in the industry, whether from their current job or previous job(s). The survey questions consist of Likert scaling questions on a scale from 1 - 5 (1 = strongly disagree, 2 = disagree, 3 = neutral, 4 = agree, and 5 = strongly agree), and brief short answers questions.

The survey has been sent to three types of industry partners. The first group is companies that develop simulation and verification tools, such as Intel and Cadence Design Systems. The second group is companies that use simulation and verification tools, such as Qualcomm and Samsung. Finally, the third group is other engineering-related companies, such as Google, META, Amazon, and Apple. Thirty-five industry professionals completed the survey.

\begin{figure}[!htbp]
    \centering
    \includegraphics[width=\textwidth]{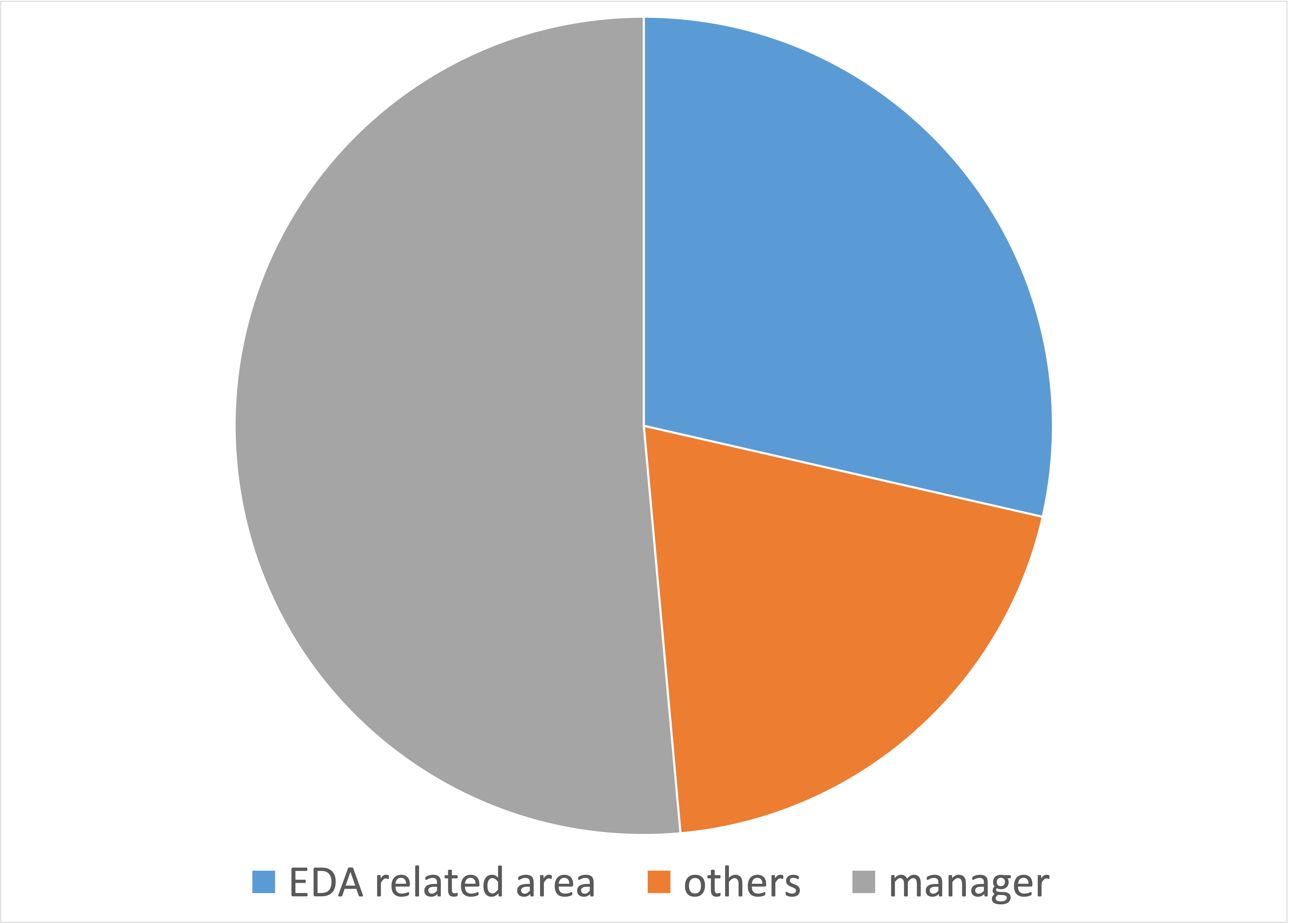}
    \caption{Current roles or job titles collected from industry professionals (N = 35).}
    \label{fig:job_titles}
\end{figure}

The survey starts with collecting job roles. The demographic results have been reported in figure \ref{fig:job_titles}. Forty-nine percent of participants are engineers who work in related areas. The remaining $51\%$ of participants are managers.

\begin{figure}[!htbp]
    \centering
    \includegraphics[width=\textwidth]{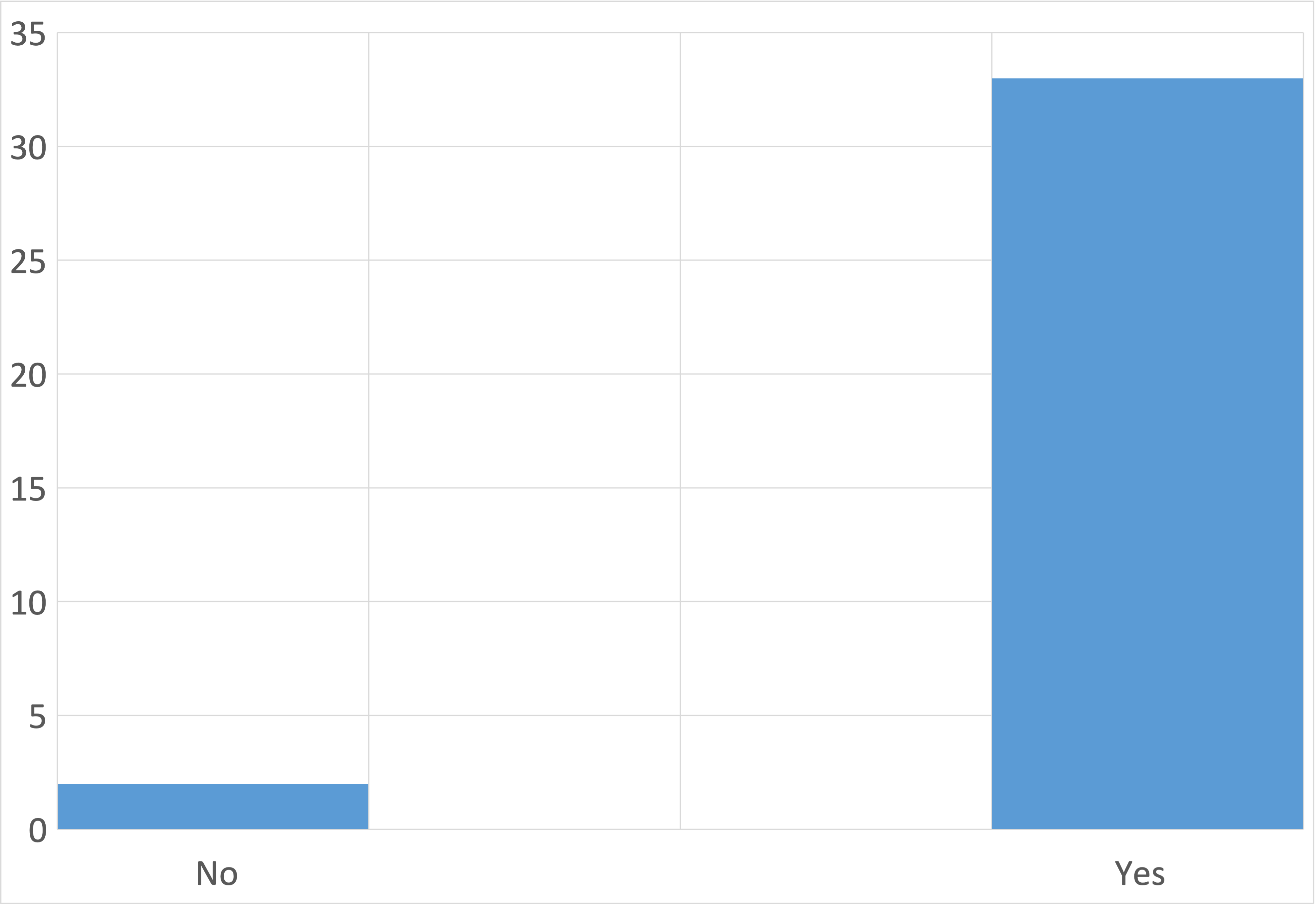}
    \caption{Check if industry professionals have used any simulation or verification tools in their job duties (N = 35).}
    \label{fig:use_simulation_before}
\end{figure}

Figure \ref{fig:use_simulation_before} shows that among thirty-five industry professionals who have filled out the survey, only two participants have not used any simulation or verification tools in their job duties. We asked industry professionals to provide examples of online experimentation environments or any setting where they are able to carry out experiments in a remote or virtual/online setting. Examples provided by participants include EDA playground, web inspector in the Chrome browser, Browserstack (a platform that allows users to access real phones), Citrix VDI, MATLAB, AWS, Google Cloud, TWCC, Xcode iPhone simulator, Labview, ORAL$\_$RT, typhoons, Bento Notebook (developed by META), and some internal tools that allow users to access FGPA, new PCB designs, and ASIC boards and servers.

Table \ref{tab:edu_survey_company} presents the perspective of industry professionals who use modeling and simulation tools to some extent in their job duties. By answering this questionnaire, industry professionals are asked to rate the statements based on their prior experiences related to simulation or verification tools. Each of these items tackles at least one objective to develop a remote laboratory for engineering education. Note that the min column is considered an outlier for this survey. Since the variance is slight, and average scores are close to each question's maximum value.

The survey results show the trend that our industry partners believe our proposed framework is well aligned with industry practices.
Based on the survey results, we conclude that new employees should be well-trained in using simulation and verification tools at school. Furthermore, a remote laboratory provides users with consistent results, easy access to analyzing and interpreting data, and developing teamwork abilities. This shows that a remote laboratory is a promising tool in engineering education that aligns with the industry's expectations of engineering graduates entering the workforce.

In summary, we interviewed students (as shown in Li \etal's work~\cite{10.1007/978-3-030-82529-4_15}) and industry professionals. The students' interviews focused on their experiences with the remote laboratory. The interviews with industry professionals zeroed in on their overall experience with simulation and verification tools. By comparing the results of the two surveys, we found that remote lab forces students to spend more time on simulation tools compared to traditional offline design. Therefore, students have considerably improved their skills in using simulation tools. This founding aligns with the study results of Hussein and Wilson's work~\cite{hussein2021}.

\begin{longtable}{|c|l|c|c|c|c|c|}
\caption{The online survey, questions, and quantitative data were collected from industry professionals (N = 35).}
\label{tab:edu_survey_company}\\
\hline
\# & \multicolumn{1}{c|}{\textbf{Question}} & \textbf{Min} & \textbf{Max} & \textbf{Mean} & \textbf{Std. Dev.} & \textbf{Var.} \\ \hline
\endhead
1 & \begin{tabular}[c]{@{}l@{}}How much do you work with \\ verification/simulation tools versus\\ real hardware in your job duties?\end{tabular} & 0 & 100\% & 0.54 & 0.33 & 0.11 \\ \hline
2 & \begin{tabular}[c]{@{}l@{}}Entry level hires are expected to\\ have solid skills in using\\ simulation and verification tools\end{tabular} & 2 & 5 & 4.06 & 1.00 & 1.00 \\ \hline
3 & \begin{tabular}[c]{@{}l@{}}Verification of hardware design is\\ typically done using simulation tools\\ prior to deploying a functional\\ solution on real hardware such as\\ FPGAs\end{tabular} & 1 & 5 & 4.06 & 1.03 & 1.06 \\ \hline
4 & \begin{tabular}[c]{@{}l@{}}Using simulation and verification tools\\ makes the design phase more efficient\end{tabular} & 3 & 5 & 4.49 & 0.66 & 0.43 \\ \hline
5 & \begin{tabular}[c]{@{}l@{}}Using an online experimentation\\ environment allows users to work\\ effectively in a team\end{tabular} & 3 & 5 & 4.11 & 0.87 & 0.75 \\ \hline
6 & \begin{tabular}[c]{@{}l@{}}Using an online experimentation\\ environment facilitates data handling.\\ For example, Amazon Web\\ Services (AWS), 1010data, etc\end{tabular} & 3 & 5 & 4.37 & 0.69 & 0.48 \\ \hline
7 & \begin{tabular}[c]{@{}l@{}}Using an online experimentation\\ environment is an adequate\\ opportunity to connect users in\\ a large variety of teams and let\\ them carry out experiments\end{tabular} & 3 & 5 & 4.21 & 0.69 & 0.47 \\ \hline
8 & \begin{tabular}[c]{@{}l@{}}Using an online experimentation\\ environment allows users to\\ obtain consistent experiences\end{tabular} & 2 & 5 & 4.29 & 0.80 & 0.64 \\ \hline
\multicolumn{1}{|l|}{9} & \begin{tabular}[c]{@{}l@{}}Feedback and assessment can be\\ made readily available by using an\\ online experimentation environment\end{tabular} & 2 & 5 & 4.09 & 0.83 & 0.69 \\ \hline
\end{longtable}

\FloatBarrier
\newpage
\section{Future Works}
\label{future_works}
The existing design of the virtual breadboard has the potential for expansion to enable students to investigate logic gates and an increased number of general-purpose inputs/outputs (GPIOs). This could lead to the development of a more advanced virtual breadboard that incorporates microcontrollers or field-programmable gate arrays (FPGAs). Although the survey data collected from industry partners is limited in scope, the participants we interviewed represented a broad range of perspectives. In the future, we plan to conduct further interviews with industry experts to arrive at more definitive conclusions about the significance of simulation and verification tools in the industry.

\FloatBarrier
\section{Conclusion}
\label{conclusion}
We utilized a remotely accessible Field Programmable Gate Arrays (FPGA) lab in teaching a junior-level course on Design of Digital Circuits and Systems. During this course, we implemented an innovative approach by incorporating a virtual breadboard that allowed students to interact with the FPGA through LEDs and switches. Typically, this type of interaction was only possible through pre-wired switches, buttons, and LEDs, such as those found on a camera. However, this approach allows for more flexibility in GPIO wiring. This is a significant advancement in providing a hybrid simulated and real-world solution for teaching GPIO labs in the classroom, giving students more autonomy than previously available. Our previous work has demonstrated the success of converting in-person lab experiments to an online format.

Furthermore, we evaluated the usefulness of simulation and experimentation tools in engineering design by conducting surveys among industry professionals and students. The results of these surveys provided insight into students' perspectives and experiences using a remote laboratory in digital systems design, as well as industry professionals' views on the importance of simulation and verification tools in engineering design. The findings indicate that remote laboratories are a promising platform for fulfilling the educational needs of students in digital design courses, while also promoting students' proficiency in using simulation and verification tools, which are highly valued by industry professionals. In the future, we plan to evaluate more assignments to provide a comprehensive statistical analysis of the benefits of using remote laboratories.

%
%
\bibliographystyle{ieeetr}
\bibliography{main}

\end{document}